\documentclass[runningheads]{acm_sen_article}
\usepackage{graphicx}
\usepackage{cite}
\usepackage{amsmath,amssymb,amsfonts}
\usepackage{textcomp}

\usepackage{graphicx}
\usepackage{color}
\usepackage{listings}
\usepackage{tikz}
\usepackage{pgfplots}
\usepackage{wrapfig}
\usepackage{mdframed}
\usepackage{enumitem}
\usepackage{url}

\usepackage{dcolumn}

\usepackage{booktabs} 
\usepackage{subfig}
\usepackage[mathscr]{euscript}

\usepackage{stmaryrd}
\usepackage{mathtools}

\usepackage{enumitem}

\DeclarePairedDelimiterX{\infdivx}[2]{(}{)}{%
  #1\;\delimsize\|\;#2%
}

\usepackage{amssymb}
\usepackage{fancyvrb}
\usepackage{graphicx}

\usepackage{tikz-qtree}

\usetikzlibrary{pgfplots.groupplots}
\usetikzlibrary{fit,shapes,arrows,automata,positioning,trees,shapes,calc}
\usetikzlibrary{plotmarks}
\usetikzlibrary{arrows.meta}

\usepackage[Procedure,ruled]{algorithm}
\usepackage[noend]{algpseudocode}

\makeatletter
\def\BState{\State\hskip-\ALG@thistlm}
\makeatother

\renewcommand{\floatpagefraction}{.95}

\usepackage{mathtools}

\usetikzlibrary{positioning} 

\tikzstyle{line}=[draw]

\tikzstyle{component} = [rectangle, rounded corners, minimum height=1cm,text centered, draw=black, fill=white!30]

\tikzstyle{input} = [rectangle, rounded corners, minimum width=3cm, minimum height=1cm,text centered, draw=white, fill=white!30]

\tikzstyle{dummy} = [rectangle, rounded corners, text centered, draw=white, fill=white!30]

\tikzstyle{arrow} = [thick,->,>=stealth]

\usepackage[nounderscore]{syntax} 

\newcommand{\mop}[1]{\mathrm{#1}} %
\newcommand*\colvec[3][]{\begin{matrix}\ifx\relax#1\relax\else#1\\\fi#2\\#3\end{matrix}}

\definecolor{mygreen}{rgb}{0,0.6,0}
\definecolor{mygray}{rgb}{0.5,0.5,0.5}
\definecolor{mymauve}{rgb}{0.58,0,0.82}

\lstdefinestyle{customc}{%
  belowcaptionskip=1\baselineskip,
  breaklines=true,
  frame=L,
  xleftmargin=\parindent,
  language=C,
  showstringspaces=false,
  basicstyle=\footnotesize\ttfamily,
  keywordstyle=\bfseries\color{green!40!black},
  commentstyle=\itshape\color{purple!40!black},
  identifierstyle=\color{blue},
  stringstyle=\color{orange},
}

\DeclareDocumentCommand \ends {} %
{%
  \mop{\it ends}
}%
\newcommand{\charat}{\mop{\it charat}} %
\tikzstyle{decision} = [diamond, draw, text width=4.5em, text badly centered,
inner sep=0pt]
\tikzstyle{block} = [rectangle, draw, fill=blue!20, text width=5em, text
centered, rounded corners, minimum height=4em]
\tikzstyle{line} = [draw, -latex']
\tikzstyle{cloud} = [draw, ellipse,fill=red!20, node distance=3cm, minimum
height=2em]
\tikzstyle{hexagon} = [regular polygon, draw, regular polygon sides=6, text
width=4.2em, inner
sep=0pt]
\tikzstyle{var} = [rectangle, draw, fill=blue!5, text centered, rounded corners,
minimum width=2em, minimum height=2em, node distance=4cm,
execute at begin node={\begin{varwidth}{12em}},
  execute at end node={\end{varwidth}}]
\tikzstyle{func} = [draw, ellipse, node distance=2cm, minimum height=2em]
\tikzstyle{sink} = [draw, ellipse, double, double distance=1pt, node
distance=2cm, minimum height=2em]

\pgfdeclarelayer{background}
\pgfdeclarelayer{foreground}
\pgfsetlayers{background,main,foreground}
\tikzstyle{materia}=[draw, fill=white!20, text width=6.0em, text centered,
minimum height=1.8em]
\tikzstyle{practica} = [materia, text width=8em, minimum width=10em,
minimum height=3em, rounded corners]
\tikzstyle{module} = [materia, text width=8em, minimum width=8em,
minimum height=3em]
\tikzstyle{driver} = [materia, text width=25em,
minimum height=2em, rounded corners]
\tikzstyle{texto} = [above, text width=6em, text centered]
\tikzstyle{linepart} = [draw, thick, color=black!50, -latex', dashed]
\tikzstyle{ur}=[draw, text centered, minimum height=0.01em]

\tikzset{
      invisible/.style={opacity=0},
      visible on/.style={alt=#1{}{invisible}},
      rect on/.style={alt=#1{rect}{}},
      alt/.code args={<#1>#2#3}{%
        \alt<#1>{\pgfkeysalso{#2}}{\pgfkeysalso{#3}} 
      },
      on slide/.code args={<#1>#2}{%
        \uncover<#1>{\pgfkeysalso{#2}}
      },
      only on slide/.code args={<#1>#2}{%
        \only<#1>{\pgfkeysalso{#2}}
      },
      rect label/.style={label={[anchor=north west, font=\tiny]north west:{\color{black!75}#1}}},
      rect/.style={rectangle, draw, rounded corners, inner sep=2pt},
      arrow/.style={edge from parent/.style={draw,-latex'}},
      left_arrow/.style={edge from parent path={(\tikzparentnode.south west) -- (\tikzchildnode.north)}},
      right_arrow/.style={edge from parent path={(\tikzparentnode.south east) -- (\tikzchildnode.north)}},
      lc_arrow/.style={edge from parent path={(\tikzparentnode.south)+(-0.5em,0pt)  -- (\tikzchildnode.north)}},
      rc_arrow/.style={edge from parent path={(\tikzparentnode.south)+(0.5em,0pt) -- (\tikzchildnode.north)}},
      down_arrow/.style={edge from parent path={(\tikzparentnode.south) -- (\tikzchildnode.north)}},
      growth parent anchor=south,
      level distance=5em,
      level 1/.style={sibling distance=12em},
      level 2/.style={sibling distance=6em}
    }

\tikzstyle{every state}=[style={font=\tiny},fill=none,draw,text=black, inner sep=0pt,minimum size=1.2em]
\tikzset{
  every edge/.append style={font=\tiny},
  every picture/.style={node distance=3em}
  }

\usepackage{multirow}
\usepackage{makecell} 
\usepackage{authblk}

\title{Attack Synthesis for Strings using Meta-Heuristics\vspace{-2ex}}
\author[1]{\small Seemanta Saha}
\author[1]{\small Ismet Burak Kadron}
\author[1]{\small William Eiers} 
\author[2]{\small Lucas Bang}
\author[1]{\small Tevfik Bultan}
\affil[1]{{\footnotesize University of California, Santa Barbara}\\
	\{seemantasaha,kadron,weiers,bultan\}@cs.ucsb.edu}
\affil[2]{{\footnotesize Harvey Mudd College}\\
 	bang@cs.hmc.edu}

\begin{document}
%
%
%

\maketitle              
%


\vspace*{-1in}

\begin{abstract}
Information leaks are a significant problem in modern computer systems
and string manipulation is prevalent in modern software.  We present
techniques for automated synthesis of side-channel attacks that recover
secret string values based on timing observations on string manipulating
code. Our attack synthesis techniques iteratively generate inputs which,
when fed to code that accesses the secret, reveal partial information about
the secret based on the timing observations, leading to recovery of the
secret at the end of the attack sequence. We use symbolic execution to
extract path constraints, automata-based model counting to estimate the
probability of execution paths, and meta-heuristic methods to maximize
information gain based on entropy for synthesizing adaptive attack steps.


\end{abstract}

\footnotetext[1]{This material is based on research supported by an Amazon Research Award and by DARPA under the agreement number FA8750-15-2-0087. The U.S. Government is 	authorized to reproduce and distribute reprints for Governmental purposes notwithstanding any copyright notation thereon. The views and conclusions contained herein are those of the authors and should not be interpreted as necessarily representing the official policies or endorsements, either expressed or implied, of DARPA or the U.S. Government.}%

\section{Introduction}
\label{sec:intro}

Modern software systems store and manipulate sensitive information. It is crucial for software developers to write code in a manner that prevents disclosure of sensitive information to arbitrary users. However, computation that accesses sensitive information can have attacker-measurable characteristics that leaks information. This can allow a malicious user to infer secret information by measuring characteristics such as execution time, memory usage, or network delay. This type of unintended leakage of secret information due to non-functional behavior of a program is called a side-channel vulnerability. In this paper, we focus on side-channel vulnerabilities that result from timing characteristics of string manipulating functions. For a given function $F$ that performs computation over strings, we automatically synthesize a side-channel attack against $F$. The synthesized attack consists of a sequence of inputs that a malicious user can use to leak information about the secret by observing timing behavior. By synthesizing an attack, we provide a proof of vulnerability for the function. 


Our approach uses symbolic execution to extract constraints characterizing the relationship between secret strings in the program, attacker controlled inputs, and side-channel observations.We compare several methods for selecting the next attack input based on meta-heuristics for maximizing the amount of information gained.




Our contributions in this paper can be summarized as follows:
(1) to the best of our knowledge, this is the first
work which performs attack synthesis specifically targeting side-channels
in string-manipulating programs; (2) we provide and experimentally compare 
several approaches to attack synthesis for strings. 
We make use of meta-heuristics for searching the
input space, including model-based searching, random searching,
simulated annealing, and genetic algorithms; and (3) we present attack synthesis 
techniques based on automata-based model counting.


\noindent \textit{\bf A Motivating Example.}
Consider a PIN-based authentication function (Fig.~\ref{fig:check1}) with 
inputs: 
1) a secret PIN $h$, and 
2) a user input, $l$.
Both $h$ and $l$ are strings of digit characters (``0''--``9'') of length $4$. 
We have adopted the nomenclature used in
security literature where $h$ denotes the {\em high-security}
value (the secret PIN) and $l$ denotes the {\em low-security} value, (the
input that the function compares with the PIN). 
The function compares the PIN and the user input
character by character and returns \texttt{false} as soon as it finds a
mismatch. Otherwise it returns \texttt{true}.

\begin{figure}[!t]
\vspace*{-0.5in}
\centering
\begin{minipage}{0.85\linewidth}
\begin{lstlisting}[stepnumber=0,basicstyle=\scriptsize]
public Boolean checkPIN(String h, String l){
  for (int i = 0; i < 4; i++)
    if (h.charAt(i) != l.charAt(i)) 
      return false;
  return true;
}
\end{lstlisting}
\end{minipage}
\caption{PIN checking example.}
\label{fig:check1}
\vspace*{-6mm}
\end{figure}

One can infer information about the secret $h$ by measuring the execution time. 
For each length of the common prefix
of $h$ and $l$, the execution time will differ. 
Notice that if $h$ and $l$ have no common prefix, then \texttt{checkPIN}
will have the shortest execution time since the loop body will be executed
only once; this corresponds to 63 Java bytecode
instructions. 
If $h$ and $l$ have a common prefix of one character, 
we see a longer execution time since the loop body executes
twice (78 instructions). 
In the case that $h$ and $l$ match completely, \texttt{checkPIN}
has the longest running time (108 instructions).
There are $5$ observable values since there are $5$ different
execution times proportional to the length of the common prefix of $h$
and $l$. Hence, an attacker can choose inputs and use
the side-channel observations to determine how much of a prefix has matched. 
For this function, we automatically generated the constraints that
characterize the length of the matching prefix and corresponding execution costs (number of executed bytecode instructions) using symbolic execution (Table~\ref{tab:example-pcs}). 
Our technique uses these constraints to synthesize an attack which
determines the value of the secret PIN. We make use of an uncertainty
function, based on Shannon entropy, to measure the progress of an attack
(Section~\ref{sec:objective_function}). Intuitively, the attacker's
uncertainty, $\mathcal{H}$ starts off at some positive value and decreases
during  the attack. When $\mathcal{H} = 0$, the attacker has fully learned
the secret (Table~\ref{tab:example-attack}).

\vspace*{-0.05in}
\begin{table}[!h]
\centering
\caption{Observation constraints generated by symbolic execution of the function in Figure~\ref{fig:check1}.}
\label{tab:example-pcs}
\begin{scriptsize}
\begin{tabular}{|l||l|l|}
\hline
$i$ & Observation Constraint, $\psi_{i}$                 & $o$       \\ \hline \hline
1   & $\charat(l,0)  \not = \charat(h,0)$         & 63      \\ \hline
2   & $\charat(l, 0)      = \charat(h, 0) \land
     \charat(l, 1) \not = \charat(h, 1)$  & 78      \\ \hline
3   & $\charat(l,0)       = \charat(h, 0) \land 
       \charat(l,1)       = \charat(h, 1) \land$  & 93      \\
    & $\charat(l,2)  \not = \charat(h, 2)$        &         \\ \hline
4   & $\charat(l,0)       = \charat(h, 0) \land
       \charat(l,1)       = \charat(h, 1) \land$   & 108     \\
    & $\charat(l,2)       = \charat(h, 2) \land
       \charat(l,3)  \not = \charat(h, 3)$   &         \\ \hline
5   & $\charat(l,0)       = \charat(h, 0) \land
         \charat(l,1)       = \charat(h, 1) \land$   & 123     \\
    & $\charat(l,2)       = \charat(h, 2) \land
        \charat(l,3)       = \charat(h, 3)$   &         \\ \hline
\end{tabular}
\end{scriptsize}
\end{table}
\vspace*{-0.05in}

\begin{table}[!h]
\centering
\caption{Attack inputs ($l$), uncertainty about the secret ($\mathcal{H}$), and observations ($o$). 
Prefix matches are shown in \textbf{bold}.}
\label{tab:example-attack}
\begin{scriptsize}
\begin{tabular}{|r|r|r|r|||r|r|r|r|}
\hline
Step & $\mathcal{H}$ & $l$ & $o$ & Step & $\mathcal{H}$ & $l$ & $o$ \\ \hline \hline
1  & 13.13 & ``\textbf{}8299'' & 63 &       15 & 5.906 & ``\textbf{13}92'' & 93  \\
2  & 12.96 & ``\textbf{}0002'' & 63 &       16 & 5.643 & ``\textbf{13}16'' & 93  \\
3  & 9.813 & ``\textbf{1}058'' & 78 &       17 & 5.321 & ``\textbf{13}08'' & 93  \\
4  & 9.643 & ``\textbf{1}477'' & 78 &       18 & 4.906 & ``\textbf{13}62'' & 93  \\
5  & 9.451 & ``\textbf{1}583'' & 78 &       19 & 4.321 & ``\textbf{13}78'' & 93  \\
6  & 9.228 & ``\textbf{1}164'' & 78 &       20 & 3.169 & ``\textbf{133}8'' & 108 \\
7  & 8.965 & ``\textbf{1}950'' & 78 &       21 & 3.000 & ``\textbf{133}2'' & 108 \\
8  & 8.643 & ``\textbf{1}220'' & 78 &       22 & 2.807 & ``\textbf{133}4'' & 108 \\
9  & 8.228 & ``\textbf{1}786'' & 78 &       23 & 2.584 & ``\textbf{133}3'' & 108 \\
10 & 7.643 & ``\textbf{1}817'' & 78 &       24 & 2.321 & ``\textbf{133}0'' & 108 \\
11 & 6.643 & ``\textbf{1}664'' & 78 &       25 & 2.000 & ``\textbf{133}5'' & 108 \\
12 & 6.491 & ``\textbf{13}42'' & 93 &       26 & 1.584 & ``\textbf{133}6'' & 108 \\
13 & 6.321 & ``\textbf{13}28'' & 93 &       27 & 0.000 & ``\textbf{1337}'' & 123 \\
14 & 6.129 & ``\textbf{13}86'' & 93 &          &       &      &   \\ \hline
\end{tabular}
\end{scriptsize}
\end{table}

Suppose that the secret is ``1337''. 
The initial uncertainty
is $\log_2 10^4 = 13.13$ bits of information.
Our attack synthesizer generated input ``8229'' at the first step and makes
an observation with cost 63, which corresponds to $\psi_1$. This
indicates that ${\charat}(h,0) \not = 8$. 
Similarly, a second synthesized input, ``0002'', implies $\charat(h,0) \not = 0$ and the uncertainty is again reduced.
At the third step the synthesized input
``1058'' yields an observation of cost 78. 
Hence, $\psi_2$ is the correct path constraint 
to update our constraints on $h$, which becomes:

\vspace*{-0.1in}

{\small $\charat(h,0) \not = 8 \land \charat(h,0) \not = 0 \land \charat(h,0) = 1 \land \charat(h,1) \not = 0$ }


\vspace*{-0.1in}

We continue synthesizing inputs and learning constraints on $h$,
which tell us more information about the prefixes of $h$, until the
secret is known after 27 steps. At the final step,
we make an observation which corresponds to $\psi_5$ indicating a full
match and the remaining uncertainty is 0. In general,
our search for attack inputs should drive the entropy to 0,
so we propose entropy optimization techniques. This particular type of attack is 
a \emph{segment attack} which is known to be a serious source of security
flaws~\cite{BAP16,CRIMEattack,Nel10}.
Our approach automatically synthesizes this attack.





\section{Automatic Attack Synthesis} 
\label{sec:attacker_model}

In this section we give a two phase approach that synthesizes attacks (Procedure~\ref{proc:attacker}). We consider functions $F$ that take as input a secret string $h \in \mathbb{H}$ and an attacker-controlled string $l \in \mathbb{L}$ and that have side-channel observations $o \in \mathbb{O}$. 

\begin{algorithm}
\begin{footnotesize}
\caption{\textsc{SynthesizeAttack}($F(h,l), C_h, h^*$) \\
This procedure calls the \textsc{GenerateConstraints} and \textsc{RunAttack} functions 
to synthesize adaptive attacks.
}
\label{proc:attacker}
\begin{algorithmic}[1]

\State $\Psi \leftarrow \textsc{GenerateConstraints}(F(h,l))$

\State \textsc{RunAttack}$(F(h,l), \Psi, C_h, h^*)$

\end{algorithmic}
\end{footnotesize}
\end{algorithm} 

\noindent \textit{\bf Static Analysis Phase.} 
The first phase generates constraints from the program for 
$h$, $l$, and $o$ (Procedure~\ref{proc:gen-constraints}).
We perform symbolic execution on the program under test with the
secret ($h$) and the attacker controlled input ($l$) marked as 
symbolic~\cite{King:1976:SEP:360248.360252,Pasareanu:2013:ASE}.
Symbolic execution runs $F$ on symbolic rather than concrete inputs
resulting in a set of path constraints $\Phi$. Each $\phi \in \Phi$ is a logical formula
that characterizes the set of inputs that execute some path in $F$.
During symbolic execution, we keep track of the side-channel observation 
for each path. 
As in other works in this area, we model the
execution time of the function by the number of instructions
executed~\cite{BAP16,Pasareanu:2016:MSC,PhanCSF2017}. We assume
that the observable values are noiseless, i.e., multiple executions of the
program with the same input value will result in the same observable value. 
We augment symbolic execution to return a function \textit{obs} that maps
a path constraint $\phi$ to an observation $o$. 
Since an attacker cannot extract information from program paths that have
indistinguishable side-channel observations, 
we combine observationally similar path constraints via disjunction (Procedure~\ref{proc:gen-constraints}, line 4), where
we say that $o \sim o'$ if $|o - o'| < \delta$ for a given threshold $\delta$. The resulting
\textit{observation constraints} (denoted $\psi_o$ and $\Psi$) characterize the relationship
between the secret ($h$) the attacker input ($l$) and side-channel observations ($o$).  

\noindent \textit{\bf Attack Synthesis Phase.} 
The second phase synthesizes a sequence of inputs that allow an attacker
to incrementally learn the secret (Procedure~\ref{proc:run-attack}). During this phase, we fix a secret
$h^*$, unknown to the attacker.
We maintain a constraint $C_h$ on the possible values of the secret $h$. 
Initially, $C_h$ merely specifies the domain of the secret.
We call procedure \textsc{AttackInput}, 
which uses one of several entropy-based heuristics (Section~\ref{sec:attack_synthesis_heuristics}),
to determine the input value $l^*$ for the current attack step. 
Then, the observation $o$ that corresponds to running the program under
attack with $h^*$ and $l^*$ is revealed.
We update $C_h$ to reflect the new constraint on $h$ implied by
the attack input and observation---we instantiate the corresponding observation constraint, $\psi_o[l \mapsto l^*]$, and conjoin it with the current $C_h$ (line 5).
Based on $C_h$, we compute an 
uncertainty measure 
for $h$ 
at every step using Shannon entropy~\cite{Cover2006}, denoted $\mathcal{H}$ 
(Section~\ref{sec:objective_function}). 
The goal is to generate inputs which drive $\mathcal{H}$ as close as possible to zero, in which case there is no uncertainty and the secret is fully known.
This attack synthesis phase repeats until it is not possible
to reduce the uncertainty, $\mathcal{H}$, any further.

\begin{algorithm}
\begin{footnotesize}
\caption{\textsc{GenerateConstraints}($F(h,l)$) \\
Performs symbolic execution on function $F$ with secret string $h$ and attacker-controlled string $l$. The resulting path constraints are combined according to indistinguishability of observations.
}
\label{proc:gen-constraints}
\begin{algorithmic}[1]


\State $\Psi \leftarrow \emptyset$

\State $(\Phi, \mathcal{O}, \textit{obs}) \leftarrow$ \textsc{SymbolicExecution}$(F(h, l))$ 

\For{$o \in \mathcal{O}$}

\State  $\psi_o \leftarrow \bigvee_{\phi \in \Phi : \textit{obs}(\phi) \sim o} \phi$

\State $\Psi \leftarrow \Psi \cup \{ \psi_o \}$

\EndFor

\State \Return $\Psi$

\end{algorithmic}
\end{footnotesize}
\end{algorithm}

\begin{algorithm}
\begin{footnotesize}
\caption{\textsc{RunAttack}($F(h,l), \Psi, C_h, h^*$) \\
Synthesizes a sequence of attack inputs, $l^*$, for $F(h,l)$, given observation constraints $\Psi$, initial constraints on $h$ ($C_h$), and unknown secret $h^*$. Function \textsc{AttackInput} is one of the variations described in Section~\ref{sec:attack_synthesis_heuristics}.
}
\label{proc:run-attack}
\begin{algorithmic}[1]

\State $\mathcal{H} \leftarrow \textsc{Entropy}(C_h)$

\While{$\mathcal{H} > 0$}

\State $l^* \leftarrow \textsc{AttackInput}(C_h, \Psi)$

\State $o \leftarrow F(h^*,l^*)$

\State $C_h \leftarrow C_h \land \phi_{o}[l \mapsto l^*]$

\State $\mathcal{H} \leftarrow \textsc{Entropy}(C_h)$


\EndWhile

\end{algorithmic}
\end{footnotesize}
\end{algorithm}

\section{Entropy-Based Objective Function } 
\label{sec:objective_function}

Here we derive an objective function to measure the amount of information
an attacker expects to gain by choosing an input value $l_{val}$ to be used in the attack
search heuristics of Section~\ref{sec:attack_synthesis_heuristics}. In the
following discussion, $H$, $L$, and $O$ are random variables representing
high-security input, low-security input, and side-channel observation
respectively. We use entropy-based metrics from the theory of quantitative
information flow~\cite{Smi09}. Given probability function
$p(h)$, 
the \emph{information entropy} of 
$H$, denoted 
$\mathcal{H}(H)$, which we interpret as the observer's \emph{uncertainty},
is 

\vspace{-5mm}

\begin{equation}
\begin{footnotesize}
\mathcal{H}(H) = - \sum_{h \in  \mathbb{H}} p(h) \log_2 {p(h)}
\label{eq:entropy}
\end{footnotesize}
\end{equation}

\vspace{-5mm}

\noindent Given conditional distributions $p(h | o, l)$, and $p(o|l)$ we quantify the attacker's expected \emph{updated uncertainty} about $h$, given a candidate choice of $L = l_{val}$, with the expectation taken over all possible observations, $o \in O$. We compute the  
\emph{conditional entropy of \ $H$ given $O$ with $L = l_{val}$} as

\vspace{-5mm}

\begin{equation}
\begin{footnotesize}
\mathcal{H}(H|O, L = l_{val}) = 
-\sum_{o \in \mathbb{O}} p(o|l_{val}) \sum_{h \in \mathbb{H}} p(h | o, l_{val}) \log_2 {p( h | o, l_{val})}
\label{eq:condentropy}
\end{footnotesize}
\end{equation}
\vspace{-5mm}

\noindent Now, we can compute the expected amount of information \textit{gained} about 
$h$ by observing
$o$ after providing input value $l_{val}$. The \emph{mutual information} between $H$ and $O$, given $L = l_{val}$ denoted $\mathcal{I}(H;O | L = l_{val})$ is the difference between the initial entropy of $H$ and the conditional entropy of $H$ given $O$ when $L = l_{val}$:

\vspace{-5mm}

\begin{equation}
\begin{footnotesize}
\mathcal{I}(H;O | L = l_{val}) = \mathcal{H}(H) - \mathcal{H}(H | O, L = l_{val})
\label{eq:mutinfo}
\end{footnotesize}
\end{equation}
\vspace{-6mm}

Equation~(\ref{eq:mutinfo}) is our objective function. Providing
input $l_{val} = l^*$ which maximizes $\mathcal{I}(H;O | L = l_{val})$
maximizes information gained about $h$. Equations~(\ref{eq:entropy})
and~(\ref{eq:condentropy}) rely on $p(h)$, $p(o|l)$, and $p(h|o,l)$, which
may change at every step of the attack. Recall that during the attack,
we maintain a constraint on the secret, $C_h$. Assuming that all secrets
that are consistent with $C_h$ are equally likely, at each step, we can
compute the required probabilities using model counting. Given a formula $F$,
performing model counting on $F$ gives the number of satisfying solutions
for $F$, which we denote $\#F$.  Thus, we observe that
$p(h) = 1 / \# C_h$ if 
$h$ satisfies $C_h$, otherwise $0$. 
Hence, Equation~(\ref{eq:entropy}) reduces to 
$\mathcal{H}(H) = \log_2(\# C_h)$.

Procedure~\ref{proc:gen-constraints} gives side-channel observations
$\mathcal{O} =$ $\{$$o_1,$ $\ldots,$ $o_n$$\}$ 
and constraints over $h$ and $l$ corresponding to each $o_i$, 
$\Psi = \{ \psi_1, \ldots, \psi_n \}$. The probability that $h$ takes on a value, constrained by a particular $\psi_i$, for a given $l_{val}$ can be computed by instantiating $\psi_i$ with $l_{val}$ and then model counting. Thus, $p(h|o_i,l_{val}) = 1 / {\# \psi_i[l \mapsto l_{val}]}$. 
Similarly, $p(o|l_{val}) = {\# \psi_i[l \mapsto l_{val}]} / \# C_h[l \mapsto l_{val}]$.

In this paper, the \textsc{Entropy} (Equation~(\ref{eq:entropy})) and
\textsc{MutualInfo} (Equation~(\ref{eq:mutinfo})) functions refer to the
appropriate entropy-based computation just described, where $p(h)$, $p(o|l)$,
and $p(h|o,l)$ are computed using the \textsc{ModelCount} procedure. We
implement the \textsc{ModelCount} procedure using the Automata-Based Model
Counter (ABC) tool, which is a constraint solver for string and integer
constraints with model counting capabilities~\cite{ABB15}. \\

%



\vspace*{-0.15in}
\section{Attack Synthesis Heuristics} 
\label{sec:attack_synthesis_heuristics}
At every attack step the goal is to choose a low input $l^*$ that 
reveals information about $h^*$. Here we describe different 
techniques for synthesizing attack inputs $l^*$. 
Each approach uses a different heuristic 
to explore a subset of the possible low inputs.
To search the input space efficiently, we first observe that we need to restrict the 
search to those $l$ that are consistent with $C_h$.





\noindent \textbf{Constraint-based Model Generation.}
Our attack synthesis algorithm maintains a constraint
$C_h$ which captures all $h$ values that are consistent with
the observations so far (Procedure~\ref{proc:run-attack}, line 5).
Using the observation constraints $\Psi$ (which identify the relation
among the secret $h$, public input $l$ and the observation $o$), we project
$C_h$ to a constraint on the input $l$, which we call $C_l$, and we restrict our search on $l$
to the set of values allowed by $C_l$.  I.e., we
only look for $l$ values that are consistent with what we know about $h$
(which is characterized by $C_h$) with respect to $\Psi$.
This approach is implemented in \textsc{GetInput} and 
\textsc{GetNeighborInput} 
functions.
To evaluate different heuristics, in our experiments we used either 
\textsc{GetInput} which returns an $l_{val}$ or \textsc{GetNeighborInput} 
which returns an $l_{val}$ by mutating the previous $l_{val}$. These two functions are 
further classified as Restricted (R), in which only models of $C_l$ are generated, or non-restricted (NR), in which we do not enforce $l_{val}$ to be a model of $C_l$. For 
Procedures~\ref{proc:random}, \ref{proc:sim-anneal},
and \ref{proc:genetic}, we can use either the restricted or non-restricted versions of 
\textsc{GetInput} and \textsc{GetNeighborInput}.



\begin{algorithm}
\begin{footnotesize}
\caption{\footnotesize \textsc{AttackInput-RA}($C_h, \Psi$) \\
Generates a low input at each attack step via random sampling.
}
\label{proc:random}
\begin{algorithmic}[1]

  \State $\mathcal{I} \leftarrow 0$

  \For{$i$ \textbf{from} $1$ \textbf{to} $K$}

  \State $l_{val} \leftarrow$ \textsc{GetInput}($\Psi, C_h$)

  \State $\mathcal{I}_{new} \leftarrow$ \textsc{MutualInfo}($\Psi, C_h, 
  l_{val}$)

  \If {$\mathcal{I}_{new} > \mathcal{I}$}

  \State $\mathcal{I} \leftarrow \mathcal{I}_{new}, l^* \leftarrow l_{val}$

  \EndIf


  \EndFor

  \State \Return $l^*$

\end{algorithmic}
\end{footnotesize}
\end{algorithm}

\noindent \textbf{Search via Random Model Generation.}
As a base-line search heuristic, we make use of the approach  described above for 
generating low values that are consistent with $C_h$. The simplest approach is to
generate a single random model from $C_l$ and use it as the next attack input.
We call this approach Model-based (M).
A slightly more sophisticated approach (Procedure~\ref{proc:random}) is to generate $K$ random samples using
$C_l$, compute the expected information gain for each of them using 
Equation~(\ref{eq:mutinfo}) 
and choose the best one.
We call this approach the Random Restricted (RA-R) heuristic (since it is restricted to
models consistent with $C_l$, and hence $C_h$).

\begin{algorithm}
  \begin{footnotesize}
    \caption{\footnotesize \textsc{AttackInput-SA}($C_h, \Psi$) \\
      Generates a low input at each attack step via simulated annealing.
    }
    \label{proc:sim-anneal}
    \begin{algorithmic}[1]
      
      \State $t \leftarrow t_0$, 
      $l_{val} \leftarrow$ \textsc{GetInput}($\Psi, C_h$),
      $\mathcal{I} \leftarrow$ \textsc{MutualInfo}($\Psi, C_h, 
      l_{val}$)
      
      \While{$t \ge t_{min}$}

      \State $l_{val} \leftarrow$ \textsc{GetNeighborInput}($l_{val},\Psi, C_h$)

      \State $\mathcal{I}_{new} \leftarrow$ \textsc{MutualInfo}($\Psi, 
      C_h, l_{val}$)

      \If {$(\mathcal{I}_{new} > \mathcal{I}) \lor \left ({e^{(\mathcal{I}_{new} -
      \mathcal{I})/t} > \textsc{RandomReal}(0,1)}\right )$}

      \State $\mathcal{I} \leftarrow \mathcal{I}_{new}, l^* \leftarrow l_{val}$

      \EndIf

      \State $t \leftarrow t - (t \times k)$

      \EndWhile

      \State \Return $l^*$

    \end{algorithmic}
  \end{footnotesize}
\end{algorithm}

\noindent \textbf{Simulated Annealing.}
Simulated annealing (SA) is a meta-heuristic for optimizing an objective
function $g(s)$~\cite{simulatedannealing}.  SA is initialized with a
candidate solution $s_0$. At step $i$, SA chooses a neighbor, $s_i$,
of candidate $s_{i-1}$. If $s_{i}$ is an improvement, i.e., $g(s_i) >
g(s_{i-1})$, then $s_i$ is used as the candidate for the next iteration. If
$s_{i}$ is not an improvement ($g(s_i) \leq  g(s_{i-1})$), then
$s_i$ is still used as the candidate for the next iteration with
a small probability $p$ calculated using the second part of disjunction at line 5 in Procedure~\ref{proc:sim-anneal}. Intuitively, SA is a controlled random search
that allows a search path to escape local optima by permitting the
search to sometimes accept worse solutions. The acceptance probability
$p$ decreases over time, which is modeled using a search
``temperature'' which ``cools off'' and converges to a steady state. Our
SA based approach is shown
in Procedure~\ref{proc:sim-anneal} where  we
use \textsc{GetNeighborInput} function to get new candidates.


\noindent \textbf{Genetic Algorithm.}
A genetic algorithm (GA) searches for an optimal input to an objective function
$g(s)$ by iteratively simulating a \textit{population} of candidate solutions
$P_i = \{s_1, \ldots s_n \}$~\cite{goldberg1989genetic}.  Each $s_i$
is modeled as a set of \textit{genes}. Here, a gene sequence
consists a string's characters.  At step $i$, we compute $g(s_j)$
as the \textit{fitness} of each candidate.  A new population $P_{i+1}$
of \textit{offspring candidates} is generated from $P_{i}$ by selecting
pairs $(s, s')$ and performing genetic crossover and mutation and 
selecting top $N$ candidates from $P_{i}$ by fitness.
Our GA-based approach is 
shown in Procedure~\ref{proc:genetic}.

Since GA applies mutation and crossover to
generate new values, restricted model generation does not 
apply. To restrict the search to
$l$ values that are consistent with $C_l$ would require implementing
mutation and crossover operations with respect to $C_l$. We are not aware
of a general approach for doing this, so during GA-based search, mutation
and crossover operations can generate low values that are inconsistent
with $C_l$ (and hence $C_h$). Such values will have no information
gain and will be ignored during search, but they can increase the search
space and slow down the search.



\begin{algorithm}[H]
  \begin{footnotesize}
    \caption{\textsc{AttackInput-GA}($C_h, \Psi$) \\
      Generates low input at each attack step using 
      a genetic algorithm. 
    }
    \label{proc:genetic}
    \begin{algorithmic}[1]
      
      \State  $\mathcal{I} \leftarrow 0$
      
      \For{$j$ \textbf{from} $1$ \textbf{to} $popSize$}

      \State $\textit{pop}[j] \leftarrow$ \textsc{GetInput}($\Psi, C_h$)

      \EndFor

      \For{$i$ \textbf{from} $1$ \textbf{to} $K$}
      
      \For{$j$ \textbf{from} $1$ \textbf{to} $\textit{popSize}$}
      
      \State $\textit{popFit}[j] \leftarrow$ \textsc{MutualInfo}($\Psi, C_h, 
      \textit{pop}[j]$)
      
      \If {$\textit{popFit}[j] > \mathcal{I}$}
      
      \State $\mathcal{I} \leftarrow \textit{popFit}[j], ~~
      l^* \leftarrow \textit{pop}[j]$
           
      \EndIf


      \EndFor

      \State $\textit{cand} \leftarrow$ \textsc{MutateAndCrossover}($\textit{pop}, \textit{popFit}, N$)

      
      \State $\textit{pop} \leftarrow$ \textsc{Append}(\textsc{SelectBest}($\textit{pop}$, $N$), $\textit{cand}$)



      
      \EndFor
      
      \State \Return $l^*$
      
    \end{algorithmic}
  \end{footnotesize}
\end{algorithm}


\section{Implementations and Experiments} 
\label{sec:experiments}


\noindent \textbf{Implementation.} 
We implemented Procedure~\ref{proc:gen-constraints} using Symbolic Path 
Finder (SPF)~\cite{Pasareanu:2013:ASE}. 
We implemented Procedure~\ref{proc:run-attack} as a Java program that takes the 
observation constraints generated by Procedure~\ref{proc:gen-constraints} as input, along with $C_h$ and $h^*$.
The variations of \textsc{AttackInput} from 
Section~\ref{sec:attack_synthesis_heuristics} 
(Procedures~\ref{proc:random}, ~\ref{proc:sim-anneal}, and~\ref{proc:genetic}) 
are implemented in Java. 
We implemented 
\textsc{GetInput}, \textsc{GetNeighborInput}, and \textsc{ModelCount} 
by extending the existing 
string model counting tool ABC. We added these features directly into 
the C++ source code of ABC along with corresponding Java APIs.

\noindent \textbf{Benchmark Details.}
To evaluate the effectiveness of our attack synthesis techniques, we experimented on a benchmark of 8 string-manipulating programs utilizing various string operations, for different string lengths (Table~\ref{tab:string-benchmark}). The functions {\tt{passCheckInsec}} and {\tt{passCheckSec}} are password checking implementations. Both compare a user input and secret password but early termination optimization (as described in the introduction) induces a timing side channel for the first one and the
latter is a constant-time implementation. We analyzed the {\tt{stringEquals}} method from the Java String library which is known to contain a timing side channel~\cite{MS14}. We discovered a similar timing side channel in {\tt{indexOf}} method from the Java String library. Function {\tt{editDistance}} example is an implementation of the standard dynamic programming algorithm to calculate minimum edit distance of two strings. Function {\tt{compress}} is a basic compression algorithm which collapses repeated substrings within two strings. {\tt{stringInequality}} and {\tt{stringCharInequality}} functions check lexicographic inequality ($<, \geq$) of two strings whereas first one directly compares the strings and later compares characters in the strings.

\renewcommand{\floatpagefraction}{.9}

\begin{table}[!h]
\centering
\caption{Benchmark details with the number of path constraints ($|\Phi|$) and the 
number of merged observation constraints ($|\Psi|$).}
\label{tab:string-benchmark}
\scalebox{0.8}{
\begin{scriptsize}
\begin{tabular}{|l|c|c|c|c|r|r|}
\hline

Benchmark & ID & Operations & {\makecell{Low\\Length}} & {\makecell{High\\Length}} & 
$|\Phi|$ 
& $|\Psi|$\\ \hline \hline

\tt{passCheckInsec} & PCI &
{\makecell{charAt,length}} & 4 & 4 & 5 & 5 \\
\hline 
\tt{passCheckSec} & PCS & {\makecell{charAt,length}} & 4 & 4 & 5 & 1 \\ \hline 

\tt{stringEquals} & SE & {\makecell{charAt,length}} & 4 & 4 & 9 & 9 
\\ 
\hline 

\tt{stringInequality} & SI & 
{\makecell{$<$,$\ge$}} & 4 & 4 & 2 & 2 \\
\hline

\tt{stringCharInequality} & SCI & {\makecell{charAt,length,$<$,$\ge$}} & 4 & 4 & 
80 & 2 \\ \hline

\tt{indexOf} & IO & {\makecell{charAt,length}} & 1 & 8 & 9 & 9 \\ \hline 

\tt{compress} & CO & {\makecell{begins,substring,length}} & 4 & 4 & 5 & 
5 \\ \hline 

\tt{editDistance} & ED & {\makecell{charAt,length}} & 4 & 4 & 2170 & 22 \\ 
\hline

\end{tabular}
\end{scriptsize}
}
\end{table}

\noindent \textbf{Experimental Setup.}
For all experiments, we use a desktop machine with an Intel Core i5-2400S 2.50 GHz CPU and 32 GB of DDR3 RAM running Ubuntu 16.04, with a Linux 4.4.0-81 64-bit kernel. We used the OpenJDK 64-bit Java VM, build 1.8.0 171. We ran each experiment for 5 randomly chosen secrets. We present the mean values of the results in Tables~\ref{tab:experiment2}. For RA, we set the sample size $K$ to 20. For SA, we set the temperature range ($t$ to $t_{min}$) from 10 to 0.001 and cooling rate $k$ as 0.1. For GA, we set population size $\textit{popSize}$ to 20, offspring size as 10, number of best selections $N$ as 10.

\noindent \textbf{Results.}
In this discussion, we describe the quality of a synthesized attack according to these metrics: the number of attack steps and overall change in entropy from $\mathcal{H}_{init}$ to $\mathcal{H}_{final}$. Attacks that do not reduce the final entropy to zero are called \textit{incomplete}.


For all benchmarks, we compare 5 approaches: 
(1) model-based (M), 
(2) non-restricted random (RA-NR),
(3) restricted random (RA-R), and
(4) restricted simulated annealing (SA R),
(5) restricted Genetic Algorithm (GA R). 
When we compare RA-NR and RA-R we observe that RA-NR is not as efficient
as reducing the entropy because attack input generation fails
to find any informative inputs for most of the steps. By restricting
the input generation to consistent models using $C_l$ as described in
Section~\ref{sec:attack_synthesis_heuristics}, we synthesize
better attacks. 
Results on non-restricted and restricted versions of SA and GA were similar.
We observe that the model-based technique (M), which also uses $C_l$ to
restrict the search space is faster than other techniques, as it greedily
uses a single random model generated by ABC as the next attack input, with no
time required to evaluate the objective function. $M$ quickly generates
attacks for most of the functions. We further examined those functions
and determined that their objective functions are ``flat'' with respect
to $l$. Any $l_{val}$ that is a model for $C_l$ at the current step yields
the same expected information gain.

\renewcommand{\floatpagefraction}{.9}

\begin{table}[htbp]
  \caption{Experimental results for model-based (M), random non-restricted (RA NR) and 
   restricted (RA R), 
   simulated-annealing restricted (SA R) and genetic algorithm restricted (GA R) heuristics. 
    Time bound is set as 3600 seconds.}
  \begin{center}
  	\scalebox{0.9}{
    \begin{scriptsize}
      \begin{tabular}{|l|r|l|r|r|r|r|r|}
        \hline
        \textbf{ID} & \textbf{$\mathcal{H}_{init}$} &
        \textbf{Metrics} & \textbf{M} & \textbf{RA NR} & \textbf{RA R} & 
        \textbf{SA R} & 
        \textbf{GA R}\\
        \hline \hline
        \multirow{3}{*}{\tt{PCI}} &
        \multirow{3}{*}{18.8} &
        Time (s) & 15.9 & 3600.0 & 3600.0 & 3600.0 & 3600.0 \\
        \cline{3-8}
        & & Steps & 54.2 & 110.0 & 39.4 & 34.5 & 41.5 \\
        \cline{3-8} 
        & & $\mathcal{H}_{final}$ & 0.0 & 9.3 & 5.7 & 8.4 & 8.5 \\
        \cline{3-8} 
        \hline \hline
        \multirow{3}{*}{\tt{PCS}} &
        \multirow{3}{*}{18.8} &
        Time (s) & 3600.0 & 3600.0 & 3600.0 & 3600.0 & 3600.0 \\
        \cline{3-8} 
        & & Steps & 118.0 & 42.5 & 41.4 & 33.2 & 38.0 \\
        \cline{3-8} 
        & & $\mathcal{H}_{final}$ & 18.8 & 18.8 & 18.8 & 18.8 & 18.8 \\
        \cline{3-8} 
        \hline \hline
        \multirow{3}{*}{\tt{SE}} &
        \multirow{3}{*}{18.8} &
        Time (s) & 22.0 & 3600.0 & 3600.0 & 3600.0 & 3600.0 \\
        \cline{3-8} 
        & & Steps & 62.2 & 85.0 & 42.6 & 25.3 & 30.8 \\
        \cline{3-8} 
        & & $\mathcal{H}_{final}$ &0.0 & 11.8 & 6.1 & 11.1 & 8.4 \\
        \cline{3-8} 
        \hline \hline 
        \multirow{3}{*}{\tt{SI}} &
        \multirow{3}{*}{18.8} &
        Time (s) & 6.1 & 3600.0 & 78.3 & 268.2 & 218.5 \\
        \cline{3-8} 
        & & Steps & 38.2 & 171.0 & 18.6 & 17.5 & 18.2 \\
        \cline{3-8} 
        & & $\mathcal{H}_{final}$ & 0.0 & 6.5 & 0.0 & 0.0 & 0.0 \\
        \cline{3-8} 
        \hline \hline
        \multirow{3}{*}{\tt{SCI}} &
        \multirow{3}{*}{18.8} &
        Time (s) & 3600.0 & 3600.0 & 3600.0 & 3600.0 & 3600.0 \\
        \cline{3-8} 
        & & Steps & 34.6 & 5.5 & 4.0 & 2.0 & 2.0 \\
        \cline{3-8} 
        & & $\mathcal{H}_{final}$ & 12.9 & 16.8 & 16.2 & 17.7 & 17.5 \\
        \cline{3-8} 
        \hline \hline
        \multirow{3}{*}{\tt{IO}} &
        \multirow{3}{*}{37.6} &
        Time (s) & 29.1 & 3600.0 & 3600.0 & 3600.0 & 3600.0 \\
        \cline{3-8} 
        & & Steps & 26.0 & 21.5 & 18.0 & 9.5 & 11.4 \\
        \cline{3-8} 
        & & $\mathcal{H}_{final}$ & 1.0 & 1.24 & 8.7 & 16.6 & 20.1 \\
        \cline{3-8} 
        \hline \hline
        \multirow{3}{*}{\tt{CO}} &
        \multirow{3}{*}{18.8} &
        Time (s) & 3600.0 & 3600.0 & 3600.0 & 3600.0 & 3600.0 \\
        \cline{3-8} 
        & & Steps & 734.0 & 183.0 & 147.0 & 83.0 & 97.8 \\
        \cline{3-8} 
        & & $\mathcal{H}_{final}$ & 13.48 & 7.9 & 9.2 & 10.3 & 9.1 \\
        \cline{3-8}
        \hline \hline
        \multirow{3}{*}{\tt{ED}} &
        \multirow{3}{*}{18.8} &
        Time (s) & 3600.0 & 3600.0 & 3600.0 & 3600.0 & 3600.0 \\
        \cline{3-8} 
        & & Steps & 27.6 & 1.0 & 1.0 & 1.0 & 1.0 \\
        \cline{3-8} 
        & & $\mathcal{H}_{final}$ & 12.6 & 18.4 & 17.8 & 17.8 & 17.8 \\
        \cline{3-8} 
        \hline
      \end{tabular}
    \end{scriptsize}
	}
    \label{tab:experiment2}
  \end{center}
\end{table}

Although M is fast and generates attacks for each benchmark, experimental
results show that it requires more attack steps compared (in terms of
information gain) to the attacks generated by meta-heuristic techniques
that optimize the objective function. As the experimental results show for
the {\tt{stringInequality}} example, a meta-heuristic technique can reduce
$\mathcal{H}_{final}$ further but with fewer attack steps compared to the
model-based approach (M). And, this case would be true for any example
where different inputs at a specific attack step have different information
gain. Our experimental results also show the differences between random search (RA) and
meta-heuristics (SA and GA). For the {\tt{stringInequality}} example,
SA is better than RA and GA. RA tries a random set of models consistent
with $C_l$ as low values, and picks the one with maximum information gain;
GA uses random models consistent with $C_l$ as the initial population
and generates more low values using mutation and crossover of characters
in the candidate strings; SA selects the first candidate as a random
model  consistent with $C_l$ and then mutates the string to get other low
values. Although GA builds the initial population using low values that are
consistent with $C_l$, mutation and crossover operations can lead to low
values which are not consistent with $C_l$. On the other hand, low values
explored by SA and RA are always consistent with $C_l$, giving better
results overall. Finally, we observe that some of our selected benchmarks
are more secure against our attack synthesizer than others. In particular,
{\tt{passCheckSec}}, a constant-time implementation of password checking,
did not leak any information through the side channel. Two other examples
from the benchmark, {\tt{stringCharInequality}} {\tt{editDistance}} also
did not succumb to our approach easily, due to the relatively large number
of generated constraints 80 and 2170 respectively, indicating a much more
complex relationship between the inputs and observations. To summarize, our
experiments indicate that our attack synthesis approach is able to construct
side-channel attacks against string manipulating programs, providing
evidence of vulnerability (e.g. {\tt{passCheckInsec}}). Further, when our
attack synthesizer fails to generate attack steps ({\tt{passCheckSec}}), or
is only able to extract a relatively small information after many steps or
significant computation time ({\tt{editDistance}}), it provides evidence that
the function under test is comparatively safer against side-channel attacks.

\vspace*{-0.05in}
\section{Related Work} 
\label{sec:related_work}

There has been prior work on analyzing side-channels~\cite{Brumley:2003:RTA:1251353.1251354,
Chen:2010:SLW:1849417.1849974, Pasareanu:2016:MSC,BAP16}.
There has been recent results on synthesizing attacks or quantifying information
leakage under a model where the attacker can make multiple runs of the
system~\cite{DBLP:conf/ccs/KopfB07,Pasareanu:2016:MSC,DBLP:conf/esorics/ChothiaKN14,BAP16,BRB18}. For example,
LeakWatch~\cite{DBLP:conf/esorics/ChothiaKN14} estimates leakage in Java
programs based on sampling program executions on concrete inputs and K\"{o}pf
et. al. \cite{DBLP:conf/ccs/KopfB07} give a multi-run analysis based on
an enumerative algorithm. There has also been prior work on quantifying information leakage
using symbolic execution and model-counting techniques for integer
constraints~~\cite{PhanCSF2017,Pasareanu:2016:MSC,BRB18}. 
There are two previous results
closely related to our work. The first~\cite{BAP16} focuses on quantifying
information flow through side channels for string-manipulating programs,
applies only for programs that have a particular form of vulnerability known
as segment oracle side-channels, and quantifies the amount of information
leakage (does not synthesize attacks).  The second~\cite{PhanCSF2017}
synthesizes side-channel attacks using either entropy-based or SAT-based
objective functions, but works only for constraints in the theories of
integer arithmetic or bit-vectors using model counters and constraint
solvers for those theories.



\vspace*{-0.05in}
\section{Conclusion}
\label{sec:conclusion}
In this paper we presented techniques for synthesizing adaptive attacks
for string manipulating programs. To the best of our knowledge this
is the first work which is able to automatically discover side channel
vulnerabilities by synthesizing attacks targeting string manipulating
functions.  We presented several heuristics for attack synthesis and
extended an existing automata-based model counter for attack synthesis.
We experimentally demonstrated the effectiveness of our approach and
compared several variations of attack-input selection heuristics.






\vspace*{0.2in}
\begin{footnotesize}
\bibliographystyle{plain}
\bibliography{biblio}
\end{footnotesize}

\end{document}